\documentclass[prx,reprint,twocolumn,amsmath,longbibliography]{revtex4-1}
\usepackage{slashbox}
\usepackage{amsmath,amssymb,bm,mathrsfs,graphicx, braket, times,amsthm,enumerate, scrextend}
\usepackage[colorlinks=true,citecolor=blue,linkcolor=blue,urlcolor=blue]{hyperref}
\usepackage{longtable}
\usepackage{multirow}
\usepackage{array}
\usepackage{bigstrut}
\usepackage[all,cmtip]{xy}
\usepackage[normalem]{ulem}
\usepackage[usenames,dvipsnames]{color}
\usepackage{makecell}
\usepackage{xcolor}
\usepackage{float}
\usepackage{ulem}
\usepackage{graphicx}

\begin{document}
\title{Magnetic hourglass fermions: from exhaustive symmetry conditions to high-throughput materials predictions}
\author{Yating Hu}
\author{Xiangang Wan}
\author{Feng Tang}\email{fengtang@nju.edu.cn}
\affiliation{National Laboratory of Solid State Microstructures and School of Physics, Nanjing University, Nanjing 210093, China and Collaborative Innovation Center of Advanced Microstructures, Nanjing University, Nanjing 210093, China}
\begin{abstract}
 Many topological band crossings (BCs) have been predicted efficiently utilizing the symmetry properties of wave-functions at high-symmetry points. Among various BCs, the so-called hourglass BCs (with the low-energy excitations dubbed as hourglass fermions) are fascinating since they can be guaranteed to exist under specific symmetry conditions even without realistic calculations. Such novel property renders the theoretical prediction on magnetic topological metals with hourglass BC (being Weyl point, Dirac point, lying in nodal loop, and so on) independent on the calculation methods and only determined by the  symmetry of crystal and magnetic structure, namely, the magnetic space group (MSG). To date, there have no magnetic material verified with hourglass fermions. Here we first list all symmetry conditions that allow hourglass BCs in the 1651 MSGs and 528 magnetic layer groups (MLGs) with  spin-orbital coupling (SOC): Only 331 MSGs and 53 MLGs can host hourglass BCs. Among these results, the essential hourglass BCs are highlighted, whose MSGs are then applied to predict hundreds of magnetic materials from the MAGNDATA magnetic materials database and  first-principles calculations in the frame of LDA+SOC+U  verify the hourglass BCs for different values of $U$. We take CsMn$_2$F$_6$, synthesized recently with a distorted pyrochlore structure to illustrate the hourglass band structure in detail which is very clear around the Fermi level and the topologically protected surface drumhead states of (100) surface are found to spread over more than one half surface Brillouin zone and only appear in a narrow energy window ($\sim 30$ meV), which could induce intriguing stability by prominent electronic correlation. The symmetry-guaranteed existence of hourglass fermion in the predicted magnetic materials is expected to be applied in manipulating band topology by applying appropriate external fields moving the hourglass BC.

\end{abstract}

\maketitle
\date{\today}
\section{INTRODUCTION}\label{I}
The past decade and a half has witnessed the explosive growth of nonmagnetic topological materials (TMs) compared with their magnetic counterparts
\cite{RMP-Kane, RMP-Qi, RMP-Bansil, TCI-Ando, RMP-AV, RMP-Balatsky}, while the verified magnetic counterparts are scarce \cite{Tokura-review, BAB-review}, though the first magnetic topological insulator dates from the experimental observation \cite{QH-exp} and theoretical interpretation \cite{QH-TKNN} of quantum Hall effect. Nearly one decade ago, through doping the nonmagnetic topological insulator with magnetic ions, quantum anomalous Hall effect was realized \cite{Yu-Science,Xue-Science}, while the intrinsic magnetic topological materials are very rare and were only theoretically predicted and experimentally verified recently \cite{MBT-SA,MBT-PRL,MBT-CPL,MBT-NM,MBT-N,MBT-S,CoSnS-NP,Magnetic-Wely-S-1,Magnetic-Wely-S-2,Magnetic-Wely-S-3}. On the other hand, the classification of band topology in all the 1651 magnetic space groups (MSGs) in the scheme of symmetry-indicator theory \cite{Po-NC, Haruki2018} or topological quantum chemistry \cite{TQC-N, M-TQC} was obtained, which not only reveals much more diverse topological phases protected by MSG symmetry, and also lay the basis for efficient magnetic topological materials discovery since one only need to calculate the symmetry transformation properties of wave-functions at high-symmetry points \cite{N-1,N-3, N-2, Xu2020}. Once some constraint arising from the compatibility relations (CRs) on the occurrences of irreducible representations (irreps) at the high-symmetry points is not satisfied, energy bands are thus enforced to cross each other, resulting in various band nodes \cite{Weyl-Wan, SMYoung, Na3Bi, Cd3As2, Nagaosa-NC, Manes, NodalChain-N, Chen-Fang-nodal-line, Balents-nodal-line, CuPdN-1,CuPdN-2, newfermions, APL, Du-CaTe, Tang-2, Wu-nodes, Tang1651,Tang-SA, Tang-NP, HgCrSe, Slager-PRX, AP-1, AP-2, AP-3, Slager-NP, Tang-nodes, Hopf-link-1, Hopf-link-2, Hopf-link-3, Hopf-link-4, yao, wwk-nodal-surface, wwk-higherorder, yao-III, yao-IV}. Interestingly, there exists a very simple-to-use rule for topological materials prediction, both for nonmagnetic and magnetic materials: one might only need to count the electron filling and once the filling is not compatible with a band insulator, band node can thus be guaranteed to appear around the Fermi level \cite{filling-PRL-1,filling-PRL-2}. As a matter of fact, the filling constraint proves to be very useful and efficient in realistic materials predictions \cite{filling-NP, filling-PRB}, especially when the materials calculations might not be very reliable due to electron correlation, leaving the topological character of the enforced band node to be identified through further analysis. As mentioned previously, symmetry information at high-symmetry points can also indicate the  existence of band node lying in high-symmetry line/plane connecting the high-symmetry points \cite{Po-NC, Tang-NP, TQC-N, Haruki2018, M-TQC} due to the breaking of CRs and thus the topological character can be identified \cite{Ono-PRX, Wu-nodes, Tang-nodes, TNSC} when knowing how CRs are broken.  However, such identification strongly depends on the representations of wave-functions at high-symmetry points and  electron correlation  in magnetic materials might reduce the prediction reliability. The first-principles prediction of magnetic topological materials usually depends on the method of incorporating the electron correlation, which not only affects the ordering of band representations at high-symmetry points, but also even gives a magnetic structure deviating much from the realistic one giving rise to a false MSG.

\begin{figure}
 \centering
 \includegraphics[width=0.45\textwidth]{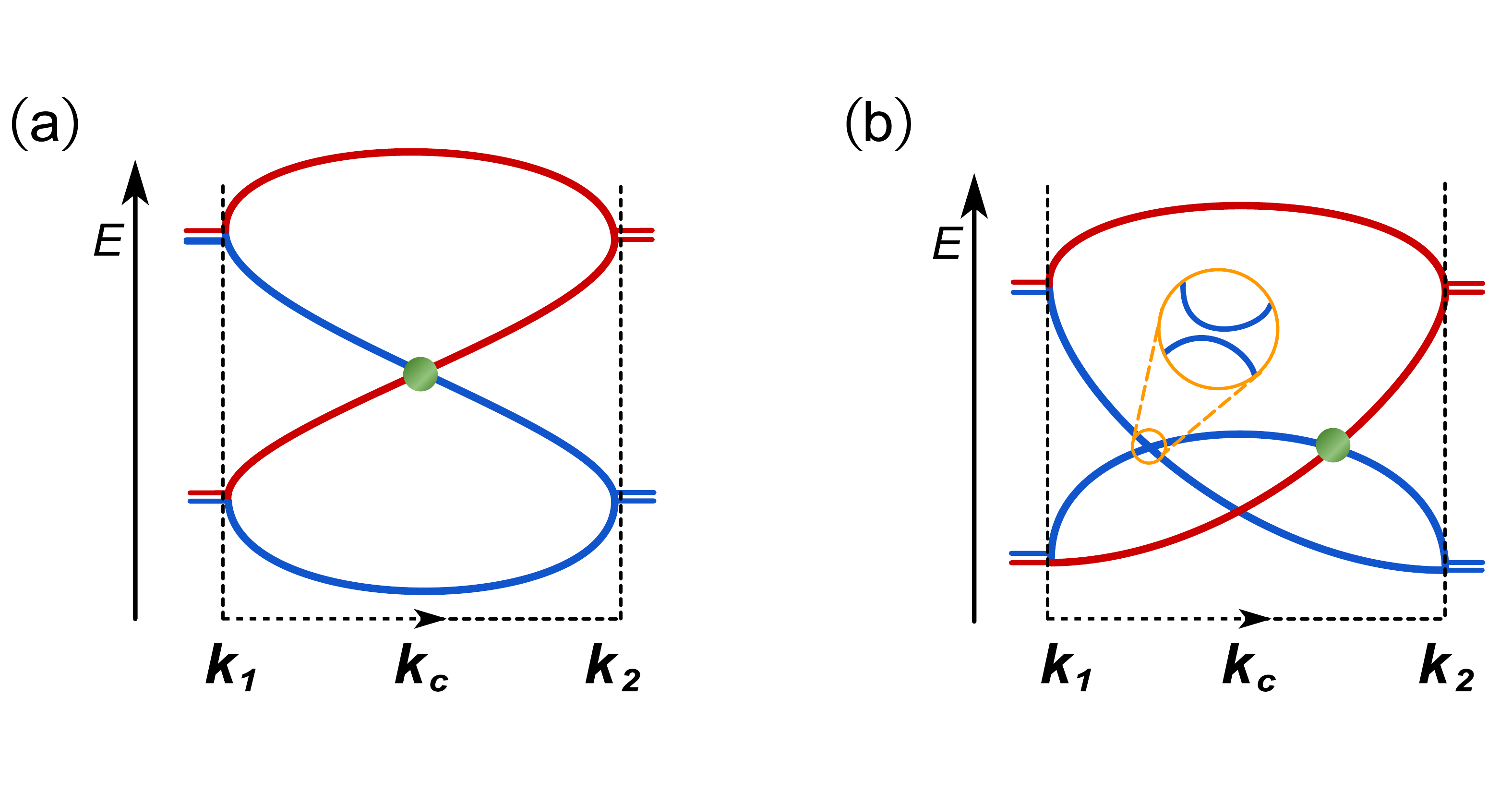}\\
 \caption{The hourglass band structure is demonstrated in (a) where the band structure is depicted along $\mathbf{k}_c$ connecting $\mathbf{k}_1$ and $\mathbf{k}_2$. The hourglass BC is denoted by a green dot which is at the crossing point of two neck bands in red and blue. Here different colors denote different (co-)irreps of the little group of $\mathbf{k}_c$. Actually the hourglass band structure here is of type A for the bottom and top bands take the same (co-)irreps as those for the neck bands (Our comprehensive study shows that type A hourglass band structure is the most common one). There are another four types of hourglass band structures described in the main text. (b) demonstrates that the realistic band structures might be twisted so that the bottom band in (a) is put upward to cross the neck bands. Interestingly, the original hourglass BC can be tuned to disappear gradually but a new hourglass BC as shown by green dot in (b) is formed.}\label{hourglass}
\end{figure}

In order to  find a simple-to-use  rule which reveals concrete topological characteristic reliably,  in this work we focus on the so-called hourglass band structure \cite{hourglass-N, ReO2, Wu-hourglass} as shown in Fig. \ref{hourglass}: When the (co-)irreps of the energy levels at $\mathbf{k}_1$ and $\mathbf{k}_2$ in the Brillouin zone (BZ) are known, the existence of band node in $\mathbf{k}_c$ which connects $\mathbf{k}_1$ and $\mathbf{k}_2$ is guaranteed. The topological characteristics of such band node can also be identified from the (co-)irreps of the band node \cite{Tang-nodes}. Furthermore, for the hourglass band structure, CRs of the 230 space groups have been applied to obtain an exhaustive list of all hourglass band structures \cite{Wu-hourglass}, indicating that in many space groups, hourglass band structure exist essentially along some high-symmetry line or plane. This means that for these space groups, we can even not perform first-principles calculations since we already know that the hourglass band structure should exist and the existence should be independent on the calculation details. On the other hand, one can expect that through applying external fields, such as electric or magnetic fields, the (co-)irreps of the energy levels at high-symmetry points retain invariant thus the band node still exists but is only movable flexibly so that large quantum responses could be anticipated (see Fig. \ref{hourglass}(b) where the hourglass band structure is twisted by external perturbations, but the hourglass band crossings (BCs) always exist). Then it is very interesting and also urgently needed to list the hourglass BCs and especially essential ones for all the 1651 MSGs, of which 1421 MSGs describe magnetically-ordered materials, which can  facilitate realistic magnetic materials predictions with topological BCs and large quantum responses. In this work,  to construct an exhaustive list of all hourglass band structures in all the 1651 MSGs and 528 magnetic layer groups (MLGs), we apply the CRs which are calculated from the (co-)irreps listed in Ref. \cite{Tang1651} of all the 1651 MSGs to obtain such list. Since we are interested in electronic materials with non-negligible spin-orbital coupling (SOC), double-valued representations  are used. Besides, for MLGs, we consider some MSGs which allow layer structure once the translation symmetry along some direction is broken. The results for MLGs can be applied to two-dimensional (2D) materials, interfaces or surfaces.

In summary, we find that there are 331 MSGs allowing hourglass BCs:  305 MSGs essentially hosting hourglass BCs among which there are 29, 73, 63 and 140 type-I, II, III and IV MSGs, respectively. We also find that there are 53 MLGs with hourglass BCs which also essentially host hourglass BCs. These MLGs could guide the synthesization of 2D hourglass semimetals and materials discovery combined with first-principles calculations \cite{WangDi-2D-2019}. Once the MSG for a given material is known, we can immediately know the position of hourglass BCs by checking the tabulation in Sec. II of Supplementary Material I.  Here for materials searches, we check the magnetic materials as listed in the MAGNDATA database \cite{MAGNDATA,MAGNDATA-web} whose magnetic structures have already been characterized. We find 175 magnetic materials which are crystallized in 48 of the 305 MSGs with essential hourglass fermions. This indicates that much more hourglass magnetic semimetals are anticipated to exist. We also perform first-principles calculations on these materials to verify the existence of hourglass BCs and select 30  materials with relatively clear hourglass band structures. Furthermore, we also find 13 magnetic materials crystallized in MSGs not in the above 305 ones but hosting hourglass BCs accidentally. These hourglass BCs can be gapped by varying the values of $U$, the parameter to incorporate the electron correlation in the first-principle calculations, as demonstrated by our calculations shown in Sec. III of Supplementary Material II. The the following, we first review briefly the strategy of obtain all symmetry conditions of the formation for hourglass BCs and highlight the essential ones. We then show three magnetic materials examples as predicted to showcase high-quality hourglass band structures near the Fermi level.

\begin{table*}
\renewcommand\arraystretch{1.3}
\caption{All MSGs and MLGs hosting essential hourglass BCs: The first column denotes the types of the MSGs and the second column contains the names of MSGs or MLGs in the Belov-Neronova-Smirnova \cite{BNS} notation. Note that when the MSG name is printed in bold, it means that the MSG can also allow an MLG hosting essential hourglass BC when the translation symmetry along some direction is broken and such direction is given in the parentheses. There are in total 305 MSGs and 53 MLGs with essential hourglass BCs. Of 305 MSGs, 288 can host type-A hourglass BCs and 94 can host type-C hourglass BCs. For MLGs,  51 and 10 can host type-A and type-C hourglass BCs, respectively. The MSGs 62.446 and 62.447 are printed in red, to which three magnetic materials are taken as examples to demonstrate hourglass BCs in the main text.}\label{essential}
\begin{tabular}{p{0.04\textwidth}<{\centering}l}
\hline\hline&
\multicolumn{1}{c}{MSG ({MLG})} \\ \hline
I & \begin{tabular}[c]{@{}l@{}}48.257; \textbf{50.277 (c)}; 52.305; \textbf{54.337 (b)}; 56.365; \textbf{57.377 (c)}; \textbf{59.405 (c)}; 60.417; 61.433; 62.441; 68.511; 70.527; 93.119; 94.127;\\98.157; 125.363; 126.375; 130.423; 132.447; 133.459; 134.471;137.507; 141.551; 142.561; 201.18;203.26; 205.33;222.98; 230.145;\end{tabular} \\ \hline
II & \begin{tabular}[c]{@{}l@{}}\textbf{4.8 (a,b)}; \textbf{7.25 (a,c)}; 9.38; \textbf{17.8 (a,b)}; \textbf{18.17 (c)}; 19.26;20.32; \textbf{26.67 (a)}; \textbf{28.88 (a,c)}; \textbf{29.100 (a)}; \textbf{30.112 (a)}; \textbf{31.124 (a)}; \textbf{32.136 (c)};\\33.145; 34.157; 36.173; \textbf{39.196 (c)};40.204; 41.212; 43.225; 45.236; 46.242; 52.306; \textbf{54.338 (b)}; 56.366; \textbf{57.378 (c)}; 60.418; 61.434;\\62.442; 76.8;77.14; 78.20; 80.30; \textbf{90.96 (c)}; 91.104; 92.112; 93.120; 94.128; 95.136; 96.144; 98.158; \textbf{100.172 (c)}; 102.188; 104.204;\\106.220; 108.234; 109.240; 110.246; \textbf{113.268 (c)}; 114.276; \textbf{117.300 (c)}; 118.308; 120.322; 122.334; 130.424; 133.460; 138.520;\\158.58; 159.62; 161.70; 169.114; 170.118; 173.130; 178.156; 179.162; 182.180; 185.198; 186.204; 188.216; 190.228; 205.34;\\219.86; 220.90;\end{tabular}\\\hline
III & \begin{tabular}[c]{@{}l@{}}\textbf{29.101 (a)}; \textbf{29.103 (a)}; 33.146; 33.148; 52.310; 52.311;52.312; \textbf{54.342 (b)}; 54.343; \textbf{54.344 (b)}; 56.369; 56.370; \textbf{57.382 (c)};\\ \textbf{57.383 (c)}; 57.384; 60.422; 60.423; 60.424; 61.436; {\color{red}62.446}; {\color{red}62.447}; 62.448; 73.551; \textbf{85.61 (c)}; 86.69; 88.83; 106.223; 110.247;\\110.249; 120.323; 124.355; \textbf{125.367 (c)}; 126.378; 126.379; \textbf{127.390 (c)}; 128.402; \textbf{129.415 (c)}; 130.426; 130.427; 130.429; 133.462;\\133.463; 133.465; 133.467; 134.475; 135.486; 135.489; 135.491; 136.498; 137.510; 137.511; 138.522; 138.523; 138.525; 141.554;\\142.564; 142.567; 219.87; 222.101; 224.113; 226.125; 227.131; 228.137;\end{tabular}\\\hline
IV & \begin{tabular}[c]{@{}l@{}}\textbf{4.10 (b)}; \textbf{7.27 (c)}; \textbf{7.29 (a)}; 7.30; 9.41; \textbf{17.11 (a)}; 17.14; \textbf{18.20 (c)}; 18.21;18.22; 19.28; 19.29; 20.36; \textbf{25.61 (a,b)}; \textbf{25.64 (a)}; 25.65;\\26.71; \textbf{26.72 (a)}; 26.76; 27.85; 27.86; \textbf{28.92 (c)}; \textbf{28.94 (a)}; \textbf{28.95 (a)}; 28.96; 28.98; 29.104; \textbf{29.105 (a)}; \textbf{29.106 (a)}; \textbf{29.107 (a)};\\29.108; 29.109; 29.110; 30.116; \textbf{30.118 (a)}; 30.120; 30.121; 30.122; 31.128; 31.129; \textbf{31.130 (a)}; 31.132; 31.133; 32.139; \textbf{32.140 (c)};\\32.142; 32.143; 33.149; 33.150; 33.151; 33.152; 33.153; 33.154;33.155; 34.160; 34.161; 34.162; 35.169; 35.171; 36.178; 38.194;\\39.201; 40.208; 41.216; 41.217; 43.228; 44.233; 45.240; 46.247; 50.286; 50.288; 52.315; 54.347; 54.351; 54.352; 56.372; 57.388;\\57.389; 59.413; 59.414; 60.426; 60.428; 60.429; 61.438; 62.452; 62.455; 77.16; 77.17; 77.18; 80.32; 90.100; 92.116; 93.124;\\93.125; 93.126; 94.132; 94.133; 94.134; 96.148; 98.162; 99.168; 99.170; 100.175; 100.178;101.184; 102.192; 102.194; 103.202;\\104.208; 105.216; 101.186; 105.217; 105.218; 106.224; 106.226; 109.244; 110.250; 111.256;111.258; 113.272; 113.274; 114.280;\\115.288; 115.290; 116.298; 117.304; 117.306; 118.312; 119.320; 122.338; 125.374; 132.458; 133.470; 134.480; 137.516;\\137.517; 138.530; 183.190; 215.73; 216.77;\end{tabular} \\ \hline
\hline
\end{tabular}
\end{table*}

\section{Brief overview of classification strategy and essential hourglass band structure}\label{II}
We firstly clarify the strategy of exhaustively listing all possible hourglass BCs based on CRs, which was firstly introduced in Ref. \cite{Wu-hourglass} and applied to the 230 space groups with and without time-reversal symmetry, corresponding to type-II and type-I MSGs, respectively. Differently, this work covers all the 1651 MSGs and adopts the convention in Ref. \cite{Bradley-Group}. Concretely, the CRs for all the MSGs are calculated firstly based on the explicit list of (co-)irreps in Ref. \cite{Tang1651}. Note that co-irreps are used when the little group contains antiunitary operation. Related with each hourglass band structure, there is one high-symmetry line or plane ($\mathbf{k}_c$ in Fig. \ref{hourglass}) connecting $\mathbf{k}_1$ and $\mathbf{k}_2$, which are high-symmetry points or lie in high-symmetry lines. Note that when $\mathbf{k}_{1/2}$  lies in a high-symmetry line, it represents an arbitrary point in the line, and in this case the infinite points in this line result in  infinite hourglass BCs  all lying in a nodal line (dubbed as hourglass nodal line). For each of $\mathbf{k}_1$ and $\mathbf{k}_2$, there are two energy levels  participating in the formation of hourglass band structure  which would split along $\mathbf{k}_c$. The band splitting pattern is encoded in the CRs. To ensure that the hourglass BC is stable, $\mathbf{k}_c$ should allow at least two different (co-)irreps and the hourglass BC is formed by the crossing of the neck bands (namely, the bands in the neck of the hourglass shape) carrying different (co-)irreps of X. Then, we can classify the hourglass band structures into five types: The bottom and top bands own different (co-)irreps and these (co-)irreps are simply those of the neck bands (type A), or only share one (co-)irrep with those of the neck bands (type B) and share no (co-)irreps with those of then neck bands (type C); The bottom and top bands own the same (co-)irrep and it take one of the (co-)irreps of the neck bands (type D) or take a different (co-)irrep from those of the neck bands (type E). For all the five types, it is easy to find that the formation of the hourglass BC  is due to the interchange of the two (co)-irreps in the hourglass BC from $\mathbf{k}_1$ to $\mathbf{k}_2$ through $\mathbf{k}_c$. We list all possible hourglass band structures in Sec. II of Supplementary Material I, for each of which we provide the coordinates of $k$ in the trio $\mathbf{k}_1-\mathbf{k}_c-\mathbf{k}_2$ following the convention in Ref. \cite{Bradley-Group} and the related CRs.

With respect to MSGs or MLGs hosting essential hourglass BC, namely, the hourglass BC definitely exists as long as the material is crystallized in the MSG or MLG, we only need to focus on type A-C hourglass band structure, since type-D and E hourglass band structures can be tuned to disappear by changing the ordering of energy levels at $\mathbf{k}_1$ or $\mathbf{k}_2$. For the type A-C hourglass band structures to be essential, it should be required that, all CRs from $\mathbf{k}_1$ (and $\mathbf{k}_2$) to $\mathbf{k}_c$ should be in the form of $D(\mathbf{k}_{1,2})^{i_{1,2}}\longrightarrow D(\mathbf{k}_c)^{j_{1,2}}\oplus D(\mathbf{k}_c)^{j'_{1,2}}$ where $i_{1,2}$ and $j_{1,2}$($j'_{1,2}$) denote the (co-)irrep of little group of $\mathbf{k}_{1,2}$ and $\mathbf{k}_c$, respectively (note that $j_{1,2}$ might be equal to $j'_{1,2}$). Besides, $\{D(\mathbf{k}_c)^{j_{1}},D(\mathbf{k}_c)^{j'_{1}}\}$ cannot be equal to $\{D(\mathbf{k}_c)^{j_{2}},D(\mathbf{k}_c)^{j'_{2}}\}$ otherwise the bands carrying these (co-)irreps can be connected and gapped from other band structures within $\mathbf{k}_c$, which is obviously not an hourglass band structure. Imposing the above requirements on essential hourglass band structures, we thus obtain all essential hourglass band structures, and as along as the MSG or MLG allows essential hourglass band structure along some high-symmetry line or plane, this MSG or MLG is listed in Table \ref{essential}. For the essential hourglass band structures, the types (namely, A, B and C) are printed in red in Sec. II of Supplementary Material I. The MSGs with essential hourglass fermions provide a very useful guide of finding magnetic topological materials.

\begin{table*}[htb]
\renewcommand\arraystretch{1.1}
\caption{Classification of 175 stoichiometric magnetic material based on magnetic structure (parallel or antiparallel and collinear) and selected magnetic materials with relatively clear hourglass band structure. The magnetic material is identified by its chemical formula, its MSG and the entry (in parentheses) in the MAGNDATA database, and separated by comma. For example, ``III, 62.446: MnCoGe (0.445)'' characterizes one magnetic material whose chemical formula is MnCoGe. Its MSG is 62.446 of type-III and its entry in the MAGNDATA database is 0.445.}\label{Materials}
\begin{tabular}{p{0.15\textwidth}<{\centering}p{0.15\textwidth}<{\centering}|p{0.03\textwidth}<{\centering}p{0.03\textwidth}<{\centering}p{0.03\textwidth}<{\centering}|p{0.05\textwidth}<{\centering}|p{0.57\textwidth}}
\hline\hline
\multicolumn{2}{c|}{Magnetic Ordering} & I & III & IV & Total & \multicolumn{1}{c}{Selected Materials} \\ \hline
\multicolumn{1}{c|}{\multirow{5}{*}{Collinear}} & Parallel & 0 & 6 & 0 & 6 & \begin{tabular}[c]{@{}l@{}}III, 62.446: MnCoGe (0.445), SrRuO$_3$ (0.732); III, 62.448: Tb$_3$NiGe$_2$ (0.439);\end{tabular}\\ \cline{2-7}
\multicolumn{1}{c|}{} & Anti-Parallel & 10 & 30 & 45 & 85 & \begin{tabular}[c]{@{}l@{}}IV, 7.29: Na$_2$MnF$_5$ (1.55); III, 33.146: Fe$_2$O$_3$ (0.299, 0.300), GaFeO$_3$ (0.38);\\I, 61.433: Ca$_2$RuO$_4$ (0.398), MnSe$_2$ (1.0.47); I, 62.441: CuFePO$_5$ (0.260),\\NiFePO$_5$ (0.261), Fe$_2$PO$_5$ (0.263); III, 62.448: NaOsO$_3$ (0.25), YRuO$_3$ (0.513);\\III, 125.367: CeMn$_2$Ge$_4$O$_{12}$ (0.189), ZrMn$_2$Ge$_4$O$_{12}$ (0.315); \end{tabular}\\ \hline
\multicolumn{2}{c|}{Non-Collinear} & 17 & 44 & 23 & 84 & \begin{tabular}[c]{@{}l@{}}I, 62.441: Mn$_2$GeO$_4$ (0.102), Co$_2$SiO$_4$ (0.218, 0.219), CoSO$_4$ (0.571, 0.96);\\III, 62.446: TeNiO$_3$ (0.94), NdSi (0.407); III, 62.447: CoFePO$_5$ (0.262),\\CsMn$_2$F$_6$ (0.726); III, 62.448: LaMnO$_3$ (0.1); I, 205.33: MnTe$_2$ (0.20);\\III, 224.113: USb (3.12), UO$_2$ (3.2), NpBi (3.7); \end{tabular}\\ \hline
\multicolumn{2}{c|}{Total} & 27 & 80 & 68 & 175 & \multicolumn{1}{c}{30} \\
\hline\hline
\end{tabular}
\end{table*}

\begin{figure*}[!t]
\centering
\includegraphics[width=1\textwidth]{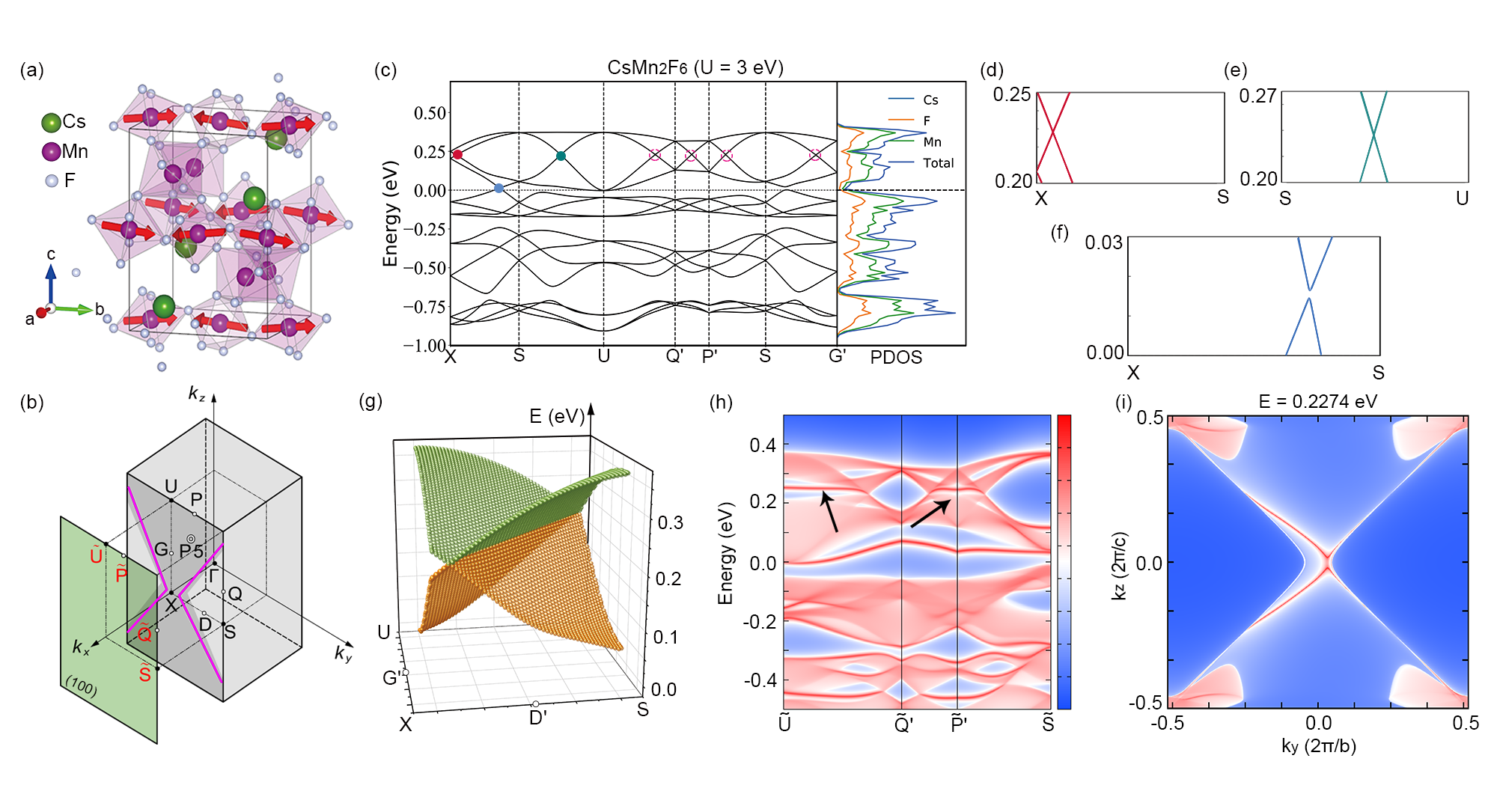}
\caption{
(a) The crystal and magnetic structure of CsMn$_2$F$_6$ crystallized in MSG 62.447. (b) The BZ of MSG 62.447 and the (100) surface BZ. (c) The electronic band structure of CsMn$_2$F$_6$ and the plot of density of states (with $U$= 3 eV). $Q', P', G'$ represents a middle point located on path $C, D, G$. The hourglass shapes are found to be very clear and also close to the Fermi level. Note that there is a tiny gap within the blue circle as shown in (f), which is consistent with the fact that the two bands in the blue circle are found to own the same co-irrep so that the hybridization between them is not prohibited. In the energy window ($-0.7$ eV,0.5 eV), we can easily find many clear hourglass band structures along different $k$ paths. We take two hourglass BCs as indicated by the red and green circles as examples and the enlarged band plots are shown in (e) and (d), respectively. (g) The 3D band plot with respect to the $k_x=\frac{\pi}{a}$ plane, namely, the plane $UXS$ shown in (b). The hourglass BCs shown in (c) with energy around 0.25 eV are all located in the nodal line in this plane, formed by the touching of two green and orange energy curves. The projection of the nodal line is shown by pink lines in (b). (h) and (i) contain the results for surface states indicated by the red color: The black arrow in (h) denotes the associated surface states while (i) is the energy contour at $0.2274$ eV for surface states. Note that the $c_{2x}$ symmetry is broken in the slab used to calculate the surface states while $c_{2z}T$ symmetry is still preserved. The drumhead surface states are found to spread over most of the surface BZ as shown by shadow area in (b), and their energies are restricted to a very narrow region ($\sim 30$ meV).}
\label{fig:2}
\end{figure*}
\begin{figure*}[!t]
\centering
\includegraphics[width=1\textwidth]{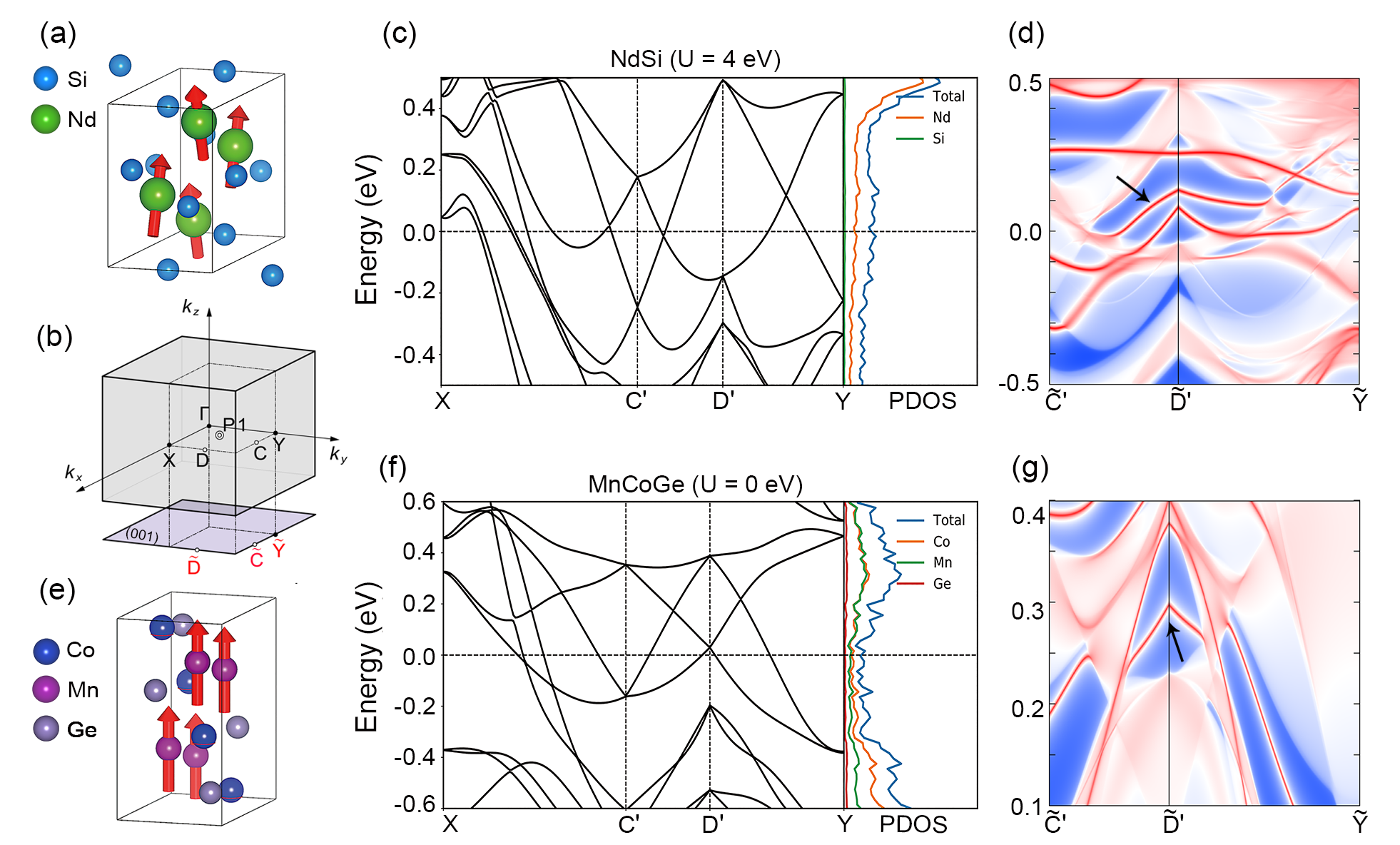}
\caption{
The results for NdSi (a,c,d) and MnCoGe (e-g). They are both crystallized in MSG 62.446 whose BZ is shown in (b) where the (001) surface BZ can also be found. (a) and (e) correspond to the crystal and magnetic structures of NdSi and MnCoGe, respectively, while (c) and (f) correspond to the electronic band structures (as well as the plots of density of states) from which the hourglass band structures can be clearly identified. For both materials, $U$ is respectively chosen to be 4 eV and 0 eV. $C',D'$ represents the middle point on path $C,D$. The hourglass BCs actually lie in a nodal line found to possess $\pi$ Berry phase, while the associated surface states for (001) surface are calculated and shown in (d) and (g), where drumhead states can be expected, indicated by black arrows.}\label{fig:3}
\end{figure*}
\section{Materials Investigation}\label{material}\label{III}
The list of all possible symmetry conditions of hourglass band structures could aid in materials predictions and design. Compared with the studies of nonmagnetic topological materials, those for the magnetic counterparts have achieved  much less progress. Recently, the magnetic topological quantum chemistry has been applied to high-throughput calculations of magnetic topological materials \cite{Xu2020}, which enlarge the candidates of magnetic topological materials, showcasing that the utilization of symmetry information can make the prediction of magnetic topological materials more efficient and especially, more reliable. However, the topological characters of the predicted symmetry-enforced magnetic topological semimetals remain unclear and the BC formed by band inversion could disappear choosing different values of $U$. With respect to the hourglass band structure here, we aim at predicting many magnetic materials hosting hourglass BCs around the Fermi level. Hence, we only need to match the MSGs of the magnetic materials candidates with the 305 ones listed in Table \ref{essential}, and due to that the hourglass band structures definitely exist along some $k$ path in these MSGs, the hourglass fermions should tend to appear very close to the Fermi level.  In total, we find 175 stoichiometric magnetic materials essentially hosting hourglass fermions as listed in the MAGNDATA database \cite{MAGNDATA,MAGNDATA-web}, belonging to 48 of the 305 MSGs. We note that the MAGNDATA database \cite{MAGNDATA,MAGNDATA-web} contains only 335 MSGs in total, indicating that more magnetic materials should be synthesized in future and thus more hourglass magnetic materials could be found.  For these materials, all materials allow the existence of type-A hourglass band structures and only 16 materials can host type-C hourglass ones. Out of these materials, we highlight 30 which are found to demonstrate relative clear hourglass band structure near the Fermi level for reasonable values of $U$ from LDA+SOC+U calculations \cite{Anisimov_1997}, as listed in Table \ref{Materials} alongside with the statistics of materials with collinear and noncollinear magnetic structures for type-I, III and IV MSGs. In the following, we showcase the hourglass band structures and detailed analysis on the topological characteristics of the hourglass BCs in magnetic materials CsMn$_2$F$_6$ \cite{CsMn2F6},  NdSi \cite{NdSi-1,NdSi-2} and MnCoGe \cite{MnCoGe-1,MnCoGe-2,MnCoGe-3}. Other than these three materials demonstrated in the main text, all the band structures of the calculated  175 magnetic hourglass materials are provided in Sec. IV of Supplementary Material II.

\subsection{Material example: CsMn$_2$F$_6$}
CsMn$_2$F$_6$ \cite{CsMn2F6}, a recently synthesized materials with distorted pyrochlore structure,  is crystallized in MSG 62.447 of type III, with orthorhombic lattice, whose BZ is shown in Fig. \ref{fig:2}(b). The unitary point group is $D_{2h}$ whose the main axis is along $x$ axis but there exists antiunitary point symmetry (namely, combination of time-reversal, $\mathcal{T}$ and spatial operation), which can be chosen as $c_{2z}\mathcal{T}$, 2-fold rotation around $z$ axis (thus $c_{2z}\mathcal{T}\cdot D_{2h}$ contains all antiunitary point symmetries). The crystal structure of this material is shown in Fig. \ref{fig:2}(a) where it can be found that Mn ions are located at the center of the octahedron of F ions. These octahedra are connected at the corner. As shown in Fig. \ref{fig:2}(a), the magnetic moments of Mn ions own  very small canting angles between each other while they are almost locked with the corresponding octahedra, indicative of significant SOC. The electronic band structure is  shown in Fig. \ref{fig:2}(c) with $U$ set to be $3$ eV.

As listed in Sec. II of Supplementary Material I, for MSG 62.447, there are high-symmetry line $D$ and high-symmetry plane $P5$ which can host hourglass band structures (see Pages 304-306 of Supplementary Material I), both of which can host type A essential hourglass band structures. The coordinates of them are $(-w,\frac{1}{2},0)$ and $(u,\frac{1}{2},v)$ in the convention adopted in Ref. \cite{Bradley-Group} (namely, in the basis of reciprocal basis lattice vectors dual to the primitive lattice basis vectors), respectively, which are $(\frac{\pi}{a},w,0)$ and $(\frac{\pi}{a},u,v)$ after transformed to the Cartesian coordinates, corresponding to $XS$ and $UXS$ as shown in Fig. \ref{fig:2}(b), respectively. Written in the trio form of $\mathbf{k}_1-\mathbf{k}_c-\mathbf{k}_2$, here we have $S$-$D$-$X$, $S$-$P5$-$X$, $X$-$P5$-$Q$, $S$-$P5$-$U$, $Q$-$P5$-$U$, $P$-$P5$-$S$, $G$-$P5$-$S$, $Q$-$P5$-$P$ and $G$-$P5$-$Q$, where $S$, $X$ and $U$ are high-symmetry points whose Cartesian coordinates are $(\frac{\pi}{a},\frac{\pi}{b},0)$, $(\frac{\pi}{a},0,0)$ and $(\frac{\pi}{a},0,\frac{\pi}{c})$, respectively and $Q$, $P$ and $G$ are high-symmetry lines whose Cartesian coordinates are $(\frac{\pi}{a},\frac{\pi}{b},w)$, $(\frac{\pi}{a},w,\frac{\pi}{b})$ and $(\frac{\pi}{a},0,w)$, respectively. Take $S-D-X$ as an example. $S$ owns two co-irreps denoted by $\{1,3\}$ and $\{2,4\}$, which would split into $\{1\}\oplus \{1\}$ and $\{2\}\oplus \{2\}$ in $D$ by CRs also shown in Sec. II of Supplementary Material I, and $X$ can only allow one co-irrep denoted by $\{1\}$ which split into $\{1\}\oplus \{2\}$ in $D$. The CRs would thus require that hourglass BC essentially appear in $D$. The hourglass band structures of the material for all the trios shown above can be easily found in Fig. \ref{fig:2}(c). Furthermore, it is easy to know that these hourglass BCs are doubly-fold-degenerate  since the two co-irreps of $D$ are both of dimension being 1.

According to the CRs-required band splitting around any band node in Ref. \cite{Tang-nodes}, we can know that the hourglass BC in $D$ actually lies in a nodal line within the high-symmetry plane $P5$. This is consistent with that $P5$ can also host hourglass BC, which definitely lies in a nodal line within $P5$. The essential existence of hourglass BCs in the above trios enforces the shape of nodal line within $P5$. As shown in Fig. \ref{fig:2}(g), the two energy branches in orange and green touch each other to form a nodal line, and through MSG symmetries, there should four nodal line segments demonstrated in Fig. \ref{fig:2}(b) by the pink lines. Note that the energy zone for this nodal line is very narrow and thus constitutes  a nearly ideal line-like Fermi surface. By Ref. \cite{Tang-nodes}  which lists all $k\cdot p$ models around all band nodes, we quickly know that  around each point in this nodal line, the low-energy $k\cdot p$ model is $(R_2q_y+R_1q_z)\sigma_0+R_3 q_x \sigma_1+R_4q_x\sigma_2+(R_6q_y+R_5q_z)\sigma_3$ where $\sigma_0$ is identity matrix, $\sigma_{x/y/z}$ is Pauli matrix, $(q_x,q_y,q_z)$ is the momentum measured from the point and $R_i$s are all real parameters. The nodal line is further found to possess $\pi$ Berry phase. Hence, drumhead surface states \cite{Balents-nodal-line} are anticipated as shown in Fig. \ref{fig:2}(h-i) and spread the shadow area in Fig. \ref{fig:2}(b). The evolution of the energy contours for (100) surface states with different energies can be found in Movie, by which the drumhead states are observed to be restricted in a very narrow energy window of around $30$ meV. Hence such flat surface states can undergo stability towards exotic ordering phase such as superconducting state by interactions \cite{Balents-nodal-line}. The interplay of magnetic order, nontrivial bulk topology, relatively clean Fermi surface and nearly perfect hourglass nodal line and exotic surface states make this material so attractive that it is expected  to be verified in future experimental studies.

\subsection{Material example: NdSi and MnCoGe}
Then we discuss materials NdSi and MnCoGe both crystallized in MSG 62.446 also of type III from Table. \ref{essential}, with orthorhombic lattice. The point group is also $D_{2h}$ but the corresponding main axis is along $z$ axis. For MSG 62.446, the antiunitary point symmetry can be chosen as $c_{2x}T$, the symmetry of all $k$ points in $P5$. Noting that $c_{2x}$ of this MSG corresponds to a screw symmetry actually, we  then find each energy level in $P5$ should be own a Kramer's degeneracy and furthermore we find that there is only one co-irrep for $P5$, thus $P5$ cannot allow hourglass band structure in this MSG, different from the above MSG. Actually, from Sec. II of Supplementary Material I, we find that only the high-symmetry plane $P1(u,v,0)$ can host essential type A hourglass BC which is further identified to lie in a nodal line within this high-symmetry plane. The related trios are found to be $X$-$P1$-$Y$, $Y$-$P1$-$D$, $X$-$P1$-$C$, $D$-$P1$-$C$ where $X$ and $Y$ are two high-symmetry points whose Cartesian coordinates are $(\frac{\pi}{a},0,0)$ and $(0,\frac{\pi}{b},0)$ and $C$ and $D$ are two high-symmetry lines whose Cartesian coordinates are $(w,\frac{\pi}{b},0)$ and $(\frac{\pi}{a},w,0)$, respectively. The electronic band structures for NdSi with $U=4$  eV and MnCoGe with $U=0$ eV are shown in Figs. \ref{fig:3}(c) and (f), demonstrating clear hourglass band structures as expected. The associated surface states are shown in Figs. \ref{fig:3}(d) and (g).

\section{CONCLUSIONS and Perspectives}\label{VI}
To conclude, we firstly obtained all possible hourglass fermions realized in 3D/2D magnetic/nonmagnetic materials purely based on symmetry arguments: Using the double-valued irrep or co-irreps  of little groups of $k$ points in the BZ and their CRs which encode the band splitting patterns, we list all concrete positions and related CRs of hourglass BCs in Supplementary Material I. In total, 331 MSGs and 53 MLGs allow existence of hourglass BCs. Furthermore, we highlight 305 MSGs essentially with hourglass BCs since any materials crystallized in any of these MSGs definitely host hourglass fermions around the Fermi level (All the 53 MLGs essentially host hourglass BCs).  232 of these 305 MSGs describe the symmetry of magnetically-ordered materials, and are around five times the number (48) of MSGs in the MAGNDATA database \cite{MAGNDATA,MAGNDATA-web}.  Then we performed high-throughput calculations based on the MSGs which essentially host hourglass BCs and demonstrate the hourglass BCs in 175 realistic magnetic materials in the MAGNDATA database \cite{MAGNDATA,MAGNDATA-web}.  The prediction of hourglass BCs in these magnetic materials is reliable since the MSGs of these materials, related with their magnetic structures, are characterized precisely \cite{MAGNDATA,MAGNDATA-web}. The magnetic structures of more magnetic materials could be characterized by advanced neutron scattering techniques, for which whether they can host hourglass fermions can be known simply by checking our tabulation.      We also expect the results of symmetry conditions could be applied in predicting more magnetic systems, such as 2D hourglass magnetic material and magnetic Hopf-link semimetals, as well as coexisting hourglass BCs with other exotic ordering such as ferroelectricity and superconductivity. The plenthora of materials predictions in this work could attract further theoretical and experimental studies in future. Finally, we expect that the MSGs/MLGs with hourglass BCs could also be utilized to inversely determine the magnetic structures of magnetic materials.
\begin{acknowledgments}
F.T. was supported by National Natural Science Foundation of China (NSFC) under Grants No. 12104215. Y.H., F.T. and X.W. were supported by the National Key R\&D Program of China (Grants No. 2018YFA0305704), NSFC Grants No. 12188101, No. 11834006, No. 51721001, and No. 11790311, and the excellent program at Nanjing University. X.W. also acknowledges the support from the Tencent Foundation through the XPLORER PRIZE.
\end{acknowledgments}

\section{Method}
The first-principles calculations, incorporating SOC for all 175  stoichiometric  magnetic materials, are implemented in the Vienna ab initio simulation package (VASP). The LDA+U scheme \cite{Anisimov_1997} was employed for different material with different Hubbard $U$ parameters: coulomb interaction $U = 0, 1, 2, 3, 4$ eV on magnetic moments of the magnetic materials with $3d/4d$ electrons, while $U = 0, 2, 4, 6$ eV for $4f/5f$ electron (refer to Sec. II of Supplementary Material II for further details). The energy cutoff was set to 500 eV and a k mesh of $0.06\pi/\mathrm{\AA}$ k-space resolution was adopted to sample the BZ. The convergence criterion for the total energy was set to be $10^{-5} \rm{eV}$. Other than calculated magnetic moments on each nonequivalent magnetic atoms, we display their energy band structures through specific high-symmetry path where hourglass fermion lies based on our symmetry analysis results in Sec. II of Supplementary  Material I. The surface states are performed on  CsMn$_2$F$_6$, NdSi and MnCoGe by $\rm{WANNIER90}$ code \cite{wannier90-1,wannier90-2} as a post processing step of the first-principles calculations. We take a disentangle calculation to construct an effective tight-binding Hamiltonian on the basis of the maximally localized Wannier function based on first-principles calculations, the method developed by Marzari and co-workers \cite{Wannier1,Wannier2}. 
\bibliography{bibtem3}

\end{document}